\documentclass[aps,twocolumn,prl,showpacs,preprintnumbers,amsmath,amssymb,superscriptaddress]{revtex4}

\usepackage{graphicx}
\usepackage{dcolumn}
\usepackage{bm}
\usepackage{hyperref}

\begin{document}

\title{Microwave Spectroscopy Evidence of Superconducting Pairing in the Magnetic-Field-Induced Metallic State of  InO$_x$ Films at Zero Temperature}

\author{Wei Liu}
\affiliation{Department of Physics and Astronomy,  Johns Hopkins University, 3400 N. Charles Street, Baltimore, MD 21218}

\author{LiDong Pan}
\affiliation{Department of Physics and Astronomy,  Johns Hopkins University, 3400 N. Charles Street,  Baltimore, MD 21218}

\author{Jiajia Wen}
\affiliation{Department of Physics and Astronomy,  Johns Hopkins University, 3400 N. Charles Street, Baltimore, MD 21218}

\author{Minsoo Kim}
\affiliation{Department of Physics, University at Buffalo-SUNY, 239 Fronczak Hall, Buffalo, NY 14260}

\author{G. Sambandamurthy}
\affiliation{Department of Physics, University at Buffalo-SUNY, 239 Fronczak Hall, Buffalo, NY 14260}

\author{N. P. Armitage }
\affiliation{Department of Physics and Astronomy,  Johns Hopkins University, 3400 N. Charles Street, Baltimore, MD 21218}

\date{\today}

\begin{abstract}
We investigate the  field tuned quantum phase transition in a 2D low-disorder amorphous InO$_x$ film in the frequency range of 0.05 to 16 GHz employing microwave spectroscopy.  In the zero temperature limit, the AC data are consistent with a scenario where this transition is from a  superconductor to a  metal instead of a direct transition to an insulator. The intervening  metallic phase is unusual with a small but finite  resistance  that is much smaller than the normal state sheet resistance at the lowest measured temperatures. Moreover, it exhibits a superconducting response on  short length and time scales while  global superconductivity is destroyed. We present evidence that the true quantum critical point of this 2D superconductor metal transition is located at a field $B_{sm}$ far below  the conventionally defined critical field $B_{cross}$ where  different isotherms of magnetoresistance cross each other. The  superfluid stiffness in the low frequency limit and the  superconducting fluctuation frequency  from opposite sides of the transition both vanish at B $\approx B_{sm}$.  The lack of  evidence for finite-frequency superfluid stiffness surviving $B_{cross}$ signifies  that $B_{cross}$ is a crossover above which superconducting fluctuations make a vanishing contribution to  DC  and AC measurements.
\end{abstract}



\maketitle
The conventional wisdom is that metallic states are prohibited at T=0 in two dimensional (2D) systems with finite disorder due to Anderson localization \cite{abrahams79}. The possible ground states for 2D disordered systems with superconducting correlations are superconductors and insulators. The 2D superconductor-insulator transition (SIT) between the two ground states is a paradigmatic example of a continuous quantum phase transition (QPT), where the transition is induced by changing a non-thermal parameter at zero temperature. It has been the subject of many theoretical and experimental studies, especially in the presence of magnetic field \cite{finkelstein,fisher90a,sondhi97a,goldman,gantmakher10a}. One possible scenario is that this transition occurs by destroying  the amplitude of the superconducting order parameter \cite{finkelstein}. Another possibility is that the Cooper pairs lose global phase coherence across the transition \cite{fisher90a}. In this case, one expects Cooper pairs to exist even in the insulator, but they are localized.

Within the bosonic description, a zero temperature metallic state can only exist at the quantum critical point (QCP) with a universal resistance of order $R_Q = h/4e^2 \approx$ 6450 $\Omega$. Although some indications for superconducting pairing in the insulator have been reported \cite{MurthyPRL04a,pourret06a,stewart07a,CranePRB07b,sacepe}, there has been little definitive evidence in favor of this pure bosonic model.  Moreover, a widely observed phenomenon that cannot be explained by either scenario is the apparent existence of 2D zero temperature metallic ground states in many experiments of disordered thin films \cite{Jaeger89a,liu92a,yazdani95a,ephron96a,Chervenak00a,okuma01a,seo06a,lin12a}, Josephson Junctions arrays \cite{JJmodel}, artificially patterned superconducting islands \cite{eley11a} and interface superconductivity \cite{reyren07a,triscone12a}.
In this metallic state, the sheet resistances ($R_{\square}$) first  drop when lowering the temperature. As T $\rightarrow$ 0, $R_{\square}$ becomes temperature independent and  saturates at a non-zero value that is  much lower than the normal state sheet resistance $R_N$, indicating the existence of a separate metallic phase in the phase diagram. This effect is usually most pronounced in low-disorder films that feature a critical sheet resistance (its value at the QCP) much lower than $R_Q$ \cite{yazdani95a,steiner}.

Despite many theoretical efforts to demonstrate the possibility of a zero-temperature dissipative state with superconducting correlations \cite{ephron96a,das99a, mason99a,kapitulnik01a, phillips02a,galitski05a}, the nature of this intermediate metallic phase is still under debate. A true metallic phase with superconducting correlations may be surprising because one might naively expect that delocalized Cooper pairs or vortices  would ultimately condense at zero temperature. On the experimental side, many groups have focused only on DC transport. AC measurements gives an advantage in studying  the 2D SIT in that one can be explicitly sensitive to temporal correlations. AC spectroscopy may reveal the true location of the QCP since it provides information about the critical slowing down of the characteristic frequency scales approaching  a transition. Through the imaginary  conductance,  microwave measurements of superconductors also allow  access to the superfluid stiffness $T_{\theta}(\omega)$, which is related to the superconducting response on a length scale set by the probing frequency.

In this Letter, we present novel measurements of frequency, temperature and field dependence of the complex microwave conductance on a  low-disorder 2D superconducting InO$_x$ film through its  QPT. Above a field $B_{sm} \approx$ 3 Tesla,  superconducting fluctuations are observed in a state with small but finite resistance as T $\rightarrow$ 0. Our main finding is that $B_{sm}$ is the true QCP for a transition from a 2D superconductor to an anomalous 2D metal at T $\rightarrow$ 0. This metallic phase is unusual due to the survival of the superconducting correlations on  short length scales at fields right above  $B_{sm}$. From the simultaneously measured DC sheet resistance $R_{\square}$, a well-defined field $B_{cross} \approx$ 7.5 Tesla is identified as the crossing point of different isotherms R(B). According to scaling theories of the resistance curves, $B_{cross}$ is conventionally interpreted to be the QCP of a 2D SIT \cite{haviland89,fisher90a,liu91a}. Contrary to the expectations for the slowing down of the fluctuations near the presumed critical field $B_{cross}$, the relevant frequency scale extrapolates to zero at the much smaller field $B_{sm}$. The superfluid stiffness $T_{\theta}$  in the zero-frequency limit vanishes from the superconducting side also at $B_{sm}$ suggesting the loss of global superconductivity near $B_{sm}$.  $T_{\theta}$  approaches zero at B $\approx B_{cross}$ in the high frequency limit indicating $B_{cross}$ only signifies a crossover to a regime where superconducting correlations are strongly suppressed even at short length scales.

Broadband microwave experiments were performed in a home-built  Corbino microwave spectrometer coupled into a He-3 cryostat \cite{liu13a, SI}. Samples are morphologically homogeneous InO$_x$ films and the nominal 2D QPT can be tuned by applying perpendicular magnetic fields \cite{Gantmakher01a,MurthyPRL04a,steiner,CranePRB07b}. We measured the complex reflectivity of the sample, from which  complex  sheet impedance and conductance can be obtained. Three calibration samples with known reflection coefficients \cite{SI} were measured to remove the contributions from the coaxial cables to the reflected signals \cite{BoothPRL96a,LeePRL01b,scheffler05b,KitanoPRB09a,liu11a,liu13a}. Two terminal DC resistance can be simultaneously  measured $via$  a bias tee.  The DC resistance without  microwave illumination was used to check and correct for any microwave induced heating \cite{SI}. Calibrations were performed at each displayed magnetic field unless otherwise specified.  With substrate corrections \cite{SI}, the true response of the InO$_x$ film can be isolated at all fields and temperatures.

\begin{figure}[h]
\begin{center}
\includegraphics[width=1\columnwidth,angle=0]{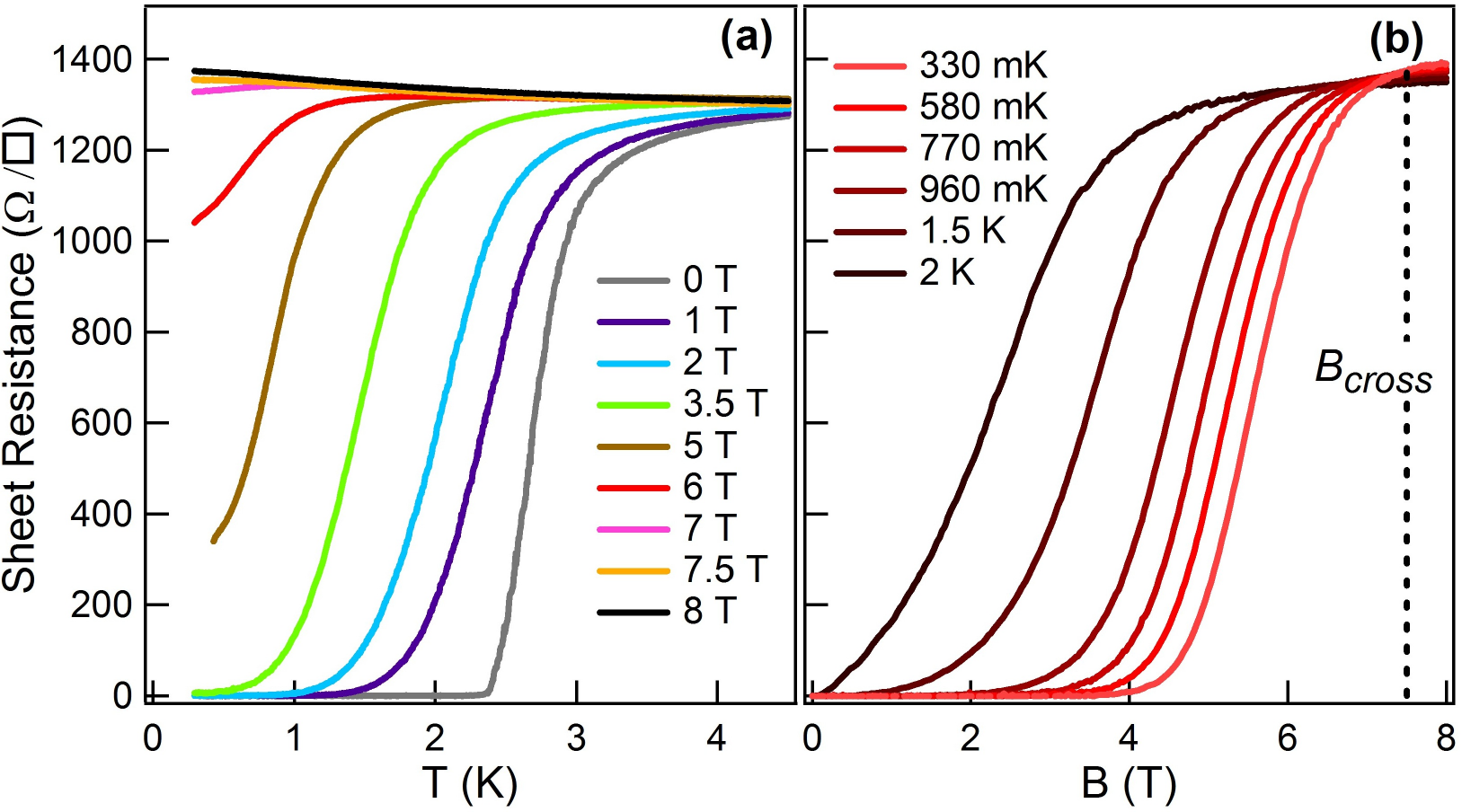}
\caption{(Color online) (a) Temperature dependence of the sheet resistance $R_{\square}$ at different fields as indicated by the color legend. (b) $R_{\square}$ as a function of field at 6 fixed temperatures as shown by the color legend. The crossing point of the two lowest temperature isotherms  is approximately  7.5 Tesla.}
\label{resistance}
\end{center}
\end{figure}

In Fig. \ref{resistance} (a) we plot the two-terminal sheet resistance $R_{\square}$ as a function of temperature at fixed magnetic fields. The InO$_x$ film studied in this paper  shows a transition to a zero resistance state at $T_c$ = 2.36 K at zero field. Previous microwave studies have demonstrated that its zero-field transition due to  thermal fluctuations are consistent with a 2D Kosterlitz-Thouless-Berezinskii type \cite{liu11a}. At low temperatures, the slopes of the resistance curves change sign at about 7.5 Tesla (see Fig. \ref{resistance} (a)).  For  the two lowest temperatures, the data also exhibit an isoresistance crossing point at 7.5 Tesla. Both observations seem to be consistent with previous DC measurements of InO$_x$ suggesting $B_{cross}$ as the QCP \cite{Murthy06a}. Comparing our resistance curves with the ones from $a$-MoGe \cite{yazdani95a} and interface superconductivity \cite{reyren07a}, we find that except for the differences in $T_c$ and $B_{cross}$, these sets of resistance curves of  three very different systems look very similar in that they all exhibit: (1) an $R_N$ that is  much smaller than $R_Q$; (2) an exceedingly weak ``insulating state", with barely a 10 \% rise in the resistance from 4 K to the lowest measured temperatures  at B $> B_{cross}$; (3)  an apparent trend towards  saturation in $R_{\square}$ toward zero temperatures for fields below $B_{cross}$. This saturation in the InO$_x$ film was confirmed  in separate two-terminal measurements down to 60 mK \cite{SI}. This implies that this InO$_x$ is very different from strongly disordered ones that show an enhancement of the resistance upwards of $10^9$ $\Omega$ with applied magnetic fields at low temperatures \cite{MurthyPRL04a,CranePRB07b}. We can characterize the  effective disorder level using  the product of Fermi wavevector ($k_F$) and electronic mean free path ($l$) \cite{SI}, which is in the range  3 $-$ 6 for this sample. It implies that this film has a much lower disorder level and falls into the same class of 2D lower-disorder superconducting thin films that usually feature a transition into a metallic phase out of the superconducting state at T $\rightarrow$ 0 \cite{steiner}.

\begin{figure}[h]
\begin{center}
\includegraphics[width=1\columnwidth,angle=0]{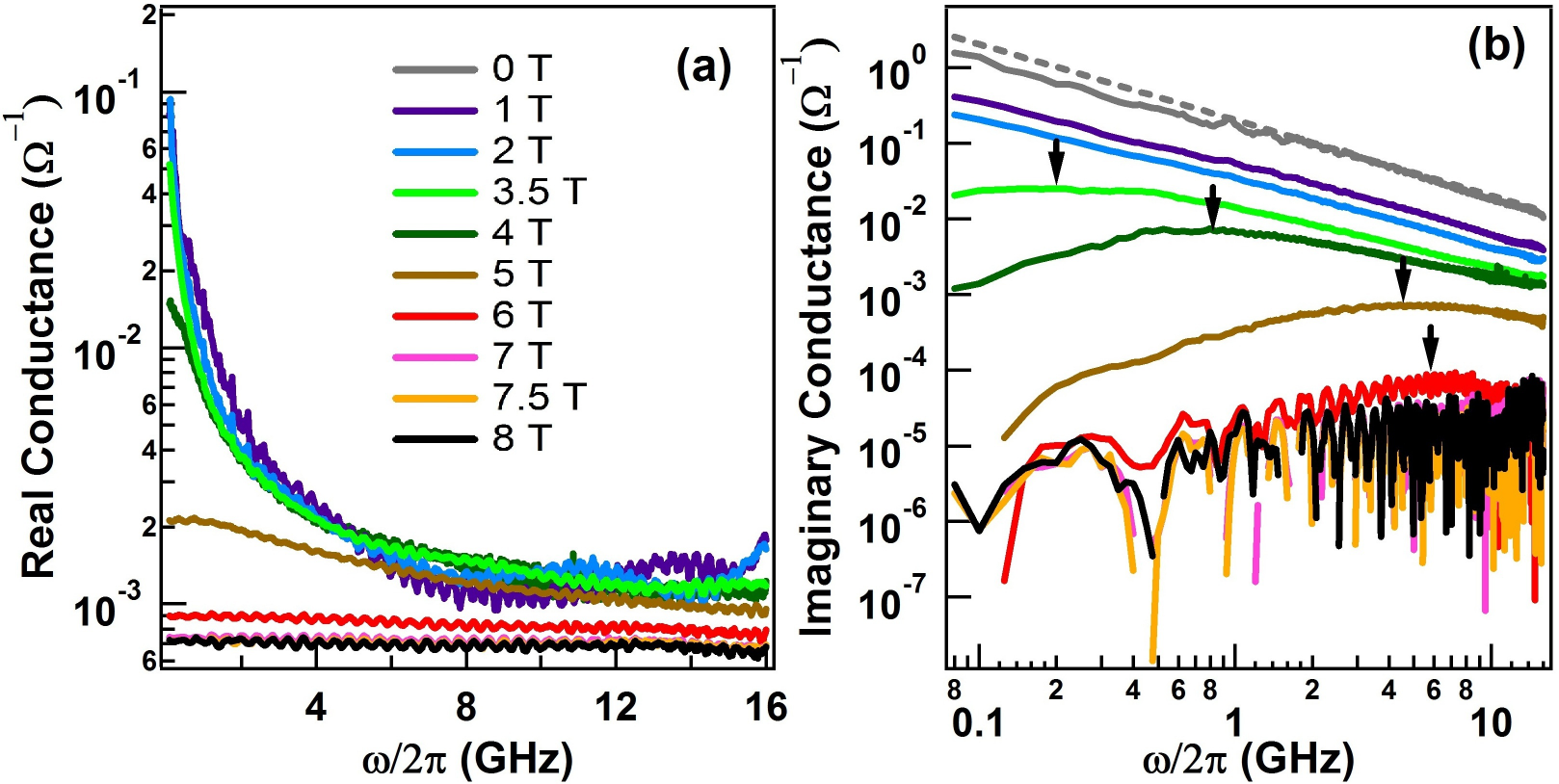}
\caption{(Color online)  Frequency dependence of the (a) real ($G_1$) and (b) imaginary ($G_2$) conductance  respectively in the ranges $\omega/2 \pi$ = 0.08 $-$ 16 GHz at the base temperature for each field. $G_1$ and $G_2$ have the same color legend at finite fields except that $G_1$ at zero field is not plotted. The dashed grey line in (b) is a guide to the eye for $G_2 \propto 1/\omega$. Arrows in (b) mark the frequencies of the maxima in $G_2$.}
\label{cond}
\end{center}
\end{figure}

In Figs. \ref{cond} (a) and  (b), we plot the real ($G_1$) and imaginary ($G_2$) conductance  as a function of frequency at the base temperatures for each field ($\approx$ 426 mK for 5 Tesla and $\approx$ 300 mK for all other fields) \cite{SI}. As shown by the straight line with a slope of -1 on the log-log plot in Fig. \ref{cond} (b), at zero field and 300 mK, $G_2$ shows the $1/\omega$ frequency dependence expected for a superconductor at frequencies below the superconducting gap ($2\Delta \approx$ 170 GHz). This dependence is consistent with  $G_1 = \frac{\pi}{2}\frac{N_se^2d}{m}\delta$($\omega$)  $via$ the Kramers-Kronig relation, where $N_s$ is the superfluid density and $d$ is the sample thickness.  Indeed, $G_1$ at zero field is small with a value that is at the limit of our experimental sensitivity \cite{liu13a,SI}.  For B $\ll B_{cross}$, $G_2$ falls as the field is applied,  but remains linear with the same slope in the log-log plot. This implies that the $\delta$-function in $G_1$ is preserved, although its spectral weight (proportional to the superfluid density) is decreasing.

At intermediate field strengths (B $\approx$ 3.5 Tesla), a maximum in $G_2$  appears (see the arrows in Fig. \ref{cond} (b)).  According to the Kramers-Kronig relation, this implies that a significant spectral component in $G_1$ has a finite width.  As shown previously \cite{liu11a}, the frequency of the maximum in $G_2$ corresponds to the characteristic fluctuation rate $\Omega$ in a fluctuating superconductor. The decrease in the frequency of the peak in $G_2$ as the field is reduced is an unambiguous signature of critical slowing down of the fluctuation frequency while approaching a continuous transition.  However, the peak in $G_2$ is developed at a field that is well below $B_{cross}$  and  the fluctuations are clearly $speeding$ $up$ as we approach $B_{cross}$ from below. This behavior is inconsistent with the conventional wisdom for QPT phenomenologies if $B_{cross}$ is a QCP, because one generally expects a $slowing$ $down$ of the fluctuation frequency scales near a continuous transition.  When B $\simeq B_{cross}$, we cannot distinguish the superconducting signal from the normal state background as $G_1$ is flat and featureless and $G_2$ is small.
\begin{figure}[h]
\begin{center}
\includegraphics[width=1\columnwidth,angle=0]{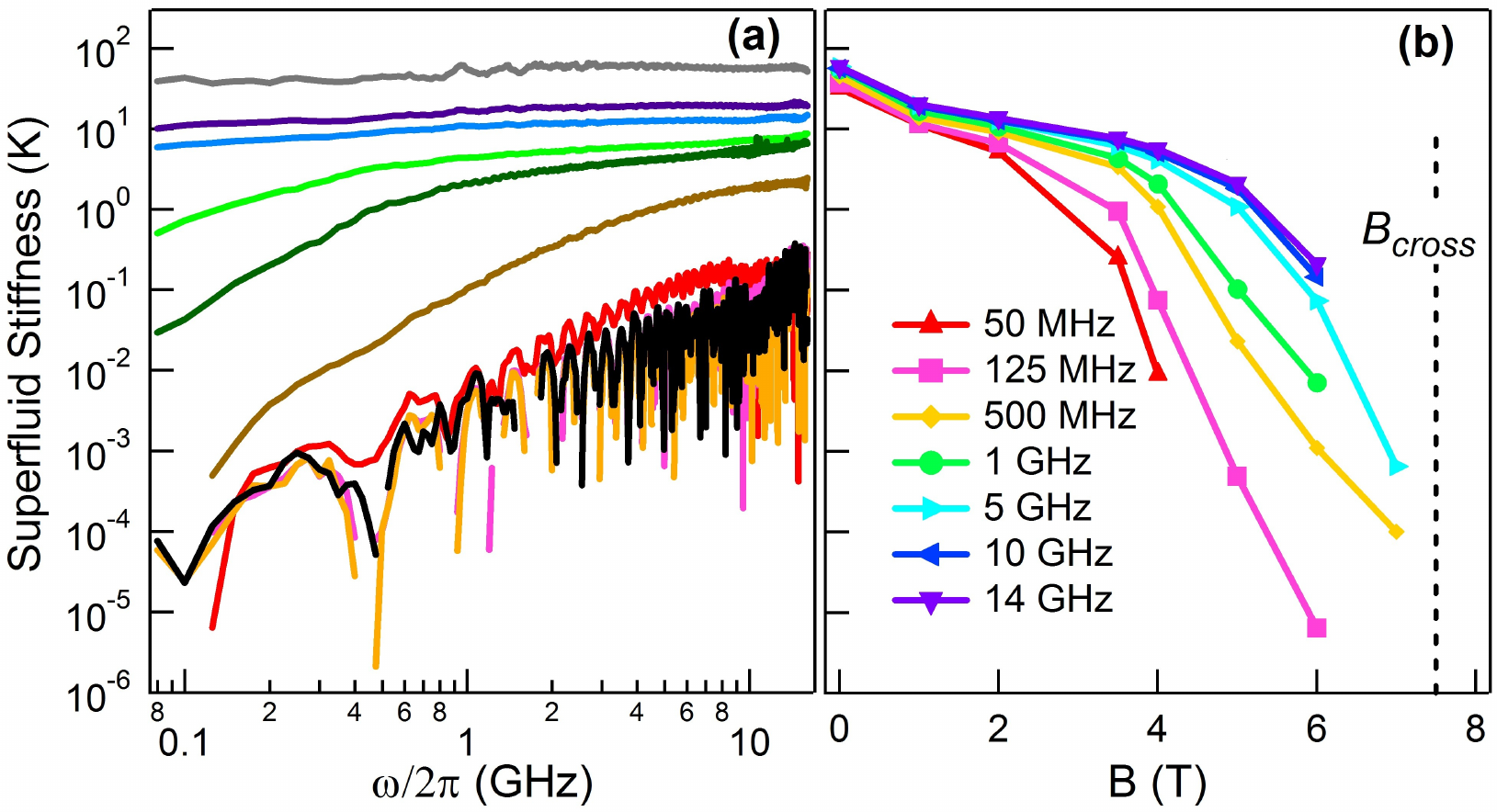}
\caption{(Color online) (a)  Frequency dependence of the superfluid stiffness in the ranges $\omega/2 \pi$ = 0.08 - 16 GHz at the base temperature for each field. The color legend for fields is the same as in Fig. \ref{cond} (a). (b)  Superfluid stiffness as a function of field at different frequencies at the base temperature for each field.}
\label{stiffness}
\end{center}
\end{figure}

An essential quantity for analyzing superconducting fluctuations is the superfluid stiffness $T_{\theta}$, which is the energy scale required to twist the phase of the superconducting order parameter. Within a parabolic band approximation, $T_{\theta} \propto N_s$, the superfluid density. More precisely (and in a model independent fashion), it is proportional to the spectral weight in  the superconducting response and can be measured through  $G_2$ as  $T_{\theta}(\omega) = \frac{G_{2}(\omega)}{G_{Q}} \frac{\hbar \omega }{k_{B}}$, where $G_{Q} = 1/R_Q$.  This relation expresses the energy scale  $T_{\theta}$ in degrees Kelvin, and gives the superfluid stiffness on a length scale set by the  probing frequency. Fig. \ref{stiffness} (a) shows  $T_{\theta} (\omega)$  at the respective base temperatures described above for each field. At zero field, $T_{\theta}$ shows essentially no frequency dependence, which suggests that the phase is ordered on all lengths. At B $\ll B_{cross}$,  $T_{\theta}$ drops but remains frequency independent. For intermediate fields, $T_{\theta}$ starts to acquire a strong frequency dependence at low $\omega$, which reflects  that Cooper pairs have short-range correlations that can be resolved at high probing frequency while the long-range correlations are suppressed. At high  $\omega$ the frequency dependence becomes less pronounced showing that one approaches a well-defined high frequency limit.

The rapid decrease in the overall scale of $T_{\theta}$ can be clearly observed in Fig. \ref{stiffness} (b) where we display field dependence of $T_{\theta}$ at several frequency cuts from Fig. \ref{stiffness} (a).  Above 2 Tesla, the curves start to spread, indicating the superconducting correlations gain a length  dependence.   At the lowest  frequency (50 MHz, which  probes the longest length scale), $T_{\theta}$ drops  around 3 Tesla indicating that long range ordered phase coherence is suppressed by increasing fields.   Note the  strong suppression in $T_{\theta}$ in this field range; at some frequencies the suppression in $T_{\theta}$ can be followed over 5 orders of magnitude.  Unlike the low frequency behavior, $T_{\theta}$ at high frequency  extrapolates towards zero near $B_{cross}$.  This latter finding differs from  previous  microwave cavity measurements  on a  more disordered InO$_x$ film \cite{CranePRB07b}. In that work the finite-frequency $T_{\theta}$ was non-zero well past the phenomenologically defined $B_{cross}$ into the strongly insulating phase. This was interpreted as an insulator with localized Cooper pairs, a state that while strongly insulating on long length scales, has superconducting correlations on short ones.  In contrast, for this low-disorder film, $T_{\theta}$ in the high frequency limit vanishes on  approaching $B_{cross}$.  This indicates that the superconducting correlations do not survive appreciably across $B_{cross}$ and the superfluid density is indistinguishable from zero into the weakly insulating state as T $\rightarrow$ 0.
\begin{figure*}[t]
\begin{center}
\includegraphics[width=2\columnwidth,angle=0]{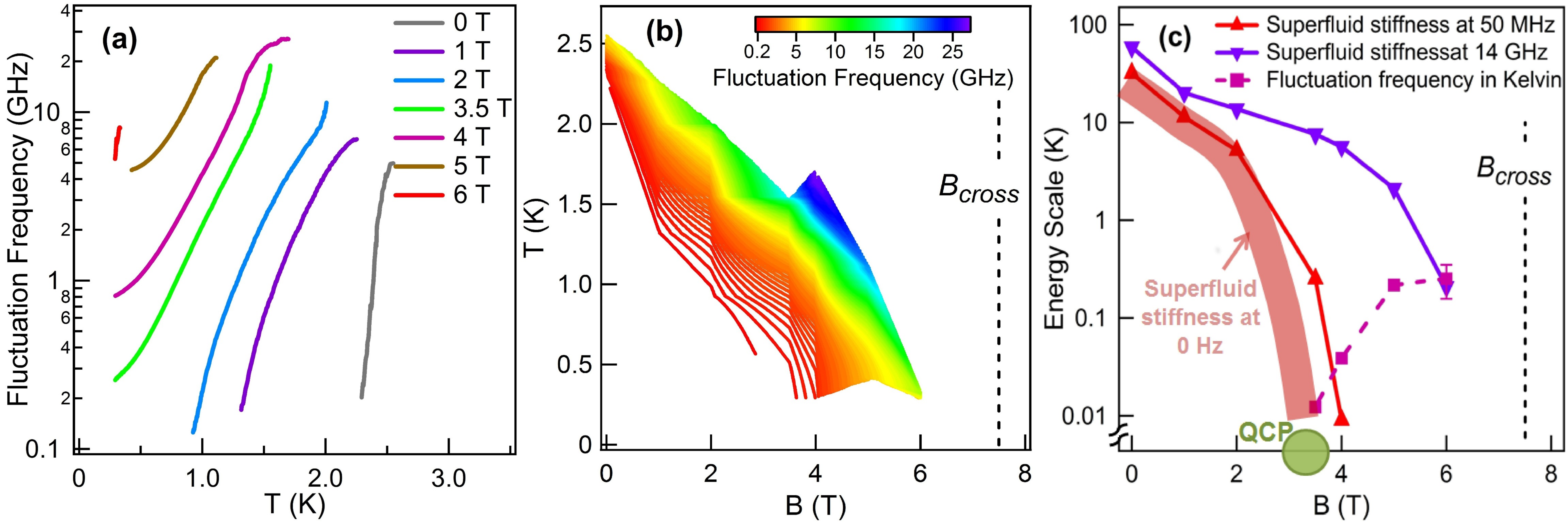}
\caption{(Color online) (a)  Temperature dependence of $\Omega$ at different fields. (b)  Contour plot of $\Omega$  in temperature and field. Color indicates the magnitude of interpolated values of $\Omega$ from the fitted data. (c)  A phase diagram of all the quantities converted to units of Kelvin. This phase diagram can be extrapolated to T = 0  since most of the quantities in the phase diagram saturate at low temperatures. The dashed vertical black lines in (b) and (c) mark $B_{cross}$.}
\label{phasediagram}
\end{center}
\end{figure*}

To form a more quantitative understanding of the fluctuations, we fit $G_1$ and $G_2$ to a model where the fluctuation contribution is given by a zero-frequency Lorentzian lineshape \cite{kuzmenko05a}.  The fitted width is the characteristic fluctuation rate $\Omega$ \cite{SI}. This is  simpler, but essentially equivalent  to the scaling analysis we performed previously \cite{liu11a}. The use of the Lorentzian lineshape is not overly restrictive and only requires the assumption that the charge currents relax exponentially in time.  The fits agree well with the  data \cite{SI}, thus justifying this assumption.

In Fig. \ref{phasediagram} (a), we plot $\Omega$(T) for fields up to 6 Tesla. Data above 6 Tesla exhibit  fluctuation rates that are far above our accessible frequency range.  At zero field, $\Omega$ goes to zero when T approaches $T_c$ from above showing the critical slowing down that confirms our previous results \cite{liu11a}. $\Omega$ drops in a much slower fashion at finite fields and even begins to saturate to a finite value as T $\rightarrow$ 0 for B  $\gtrsim$ 3.5 Tesla.  Fig. \ref{phasediagram} (b) is a contour plot of $\Omega$ in field and  temperature. Contours give lines of constant $\Omega$. It is safe to conclude that the $\Omega$ = 0 contour falls below the lowermost curve which is  $\Omega$ = 0.2 GHz. In general, small $\Omega$ contours extrapolate to zero temperature at a field  less than 4 Tesla, which is again much smaller than  $B_{cross}$.

To form a global view of the zero temperature behavior, we bring a number of quantities measured at the base temperature together in the phase diagram in Fig. \ref{phasediagram} (c). For all quantities, energy scales and frequencies have been converted to energy units (in degrees Kelvin).  Since these quantities in the phase diagram have little temperature dependence at low temperatures, this phase diagram can be extrapolated to  T = 0. In Fig. \ref{phasediagram} (c),  upward and downward triangles show the low (50 MHz) and high (14 GHz) frequency limits of $T_{\theta}$ in the accessible frequency range of our setup. The  hypothetical  behavior of  $T_{\theta}$ in the zero-frequency limit and the measured $\Omega$(T) at base temperatures (the thick line  and squares respectively in Fig. \ref{phasediagram} (c)) converge towards zero  at B $\approx $ 3 Tesla. This ``V" shaped phase diagram is exactly what one expects near a QCP where energy scales extrapolate to zero from either side.   Again, $B_{cross}$, which is conventionally considered to be a QCP, appears to be completely unrelated to the actual critical behavior.   One can see that $B_{cross}$ is the field scale where the high frequency $T_{\theta}$  is suppressed. Due to the lack of evidence for a diverging sheet resistance at $B_{sm} <$ B $< B_{cross}$ in the zero-temperature limit, one reasonable interpretation of the phase diagram is  that this low-disorder InO$_x$ film has a true QCP (the green dot in Fig. \ref{phasediagram} (c)) located at $B_{sm} \approx$ 3 Tesla between a superconducting and an anomalous 2D metallic state. The finite $T_{\theta}$ at finite frequencies indicates the existence of superconducting response on the short length scales in this metallic phase. Therefore, the 2D QPT here is characterized by the loss of global coherence in the phase of the superconducting order parameter. Dissipation may occur through quantum delocalized vortices at T = 0 \cite{ephron96a,Chervenak00a,okuma01a,seo06a,lin12a}. In this picture $B_{cross}$  only marks a crossover in behavior between a metallic state with strong superconducting correlations on short length scales and one with vanishing such correlations.

To conclude, we find evidence for a  scenario where a 2D QPT in $weakly$ disordered films occurs at a field $B_{sm}$ instead of $B_{cross}$. Although this observation runs counter to  prevailing dogma in the field, we propose  $B_{sm}$ as the QCP and a re-examination of the previous scaling analysis of transport properties in such samples. Our results have relevance to many other  systems including high temperature superconductors and interface superconductivity. The future direction of our project is to perform a careful and complete investigation of more disordered films to compare the affects of different disorder levels.

We thank S. Chakravarty, M. Feigel'man, T. Giamarchi, A. Kapitulnik, S. Kivelson, N. Markovic, Karen Michaeli, S. Sachdev, Z. Tesanovic, J.-M. Triscone and R. Valdes Aguilar for helpful discussions. The research at JHU and UB were supported by NSF DMR-0847652 and DMR-0847324 respectively.

\emph{Note added} - After completion of this work, we became aware of a new paper using two-coil mutual inductance measurements of field-tuned InO$_x$ and MoGe films \cite{misra13a}. They also observed that the true critical field is smaller than $B_{cross}$.

\newpage
\begin{center}
\large\textbf{SUPPLEMENTARY INFORMATION}
\end{center}

\section{Saturation of the resistance at the lowest measured temperatures}
\begin{figure}[thb]
\begin{center}
\includegraphics[width=\columnwidth,angle=0]{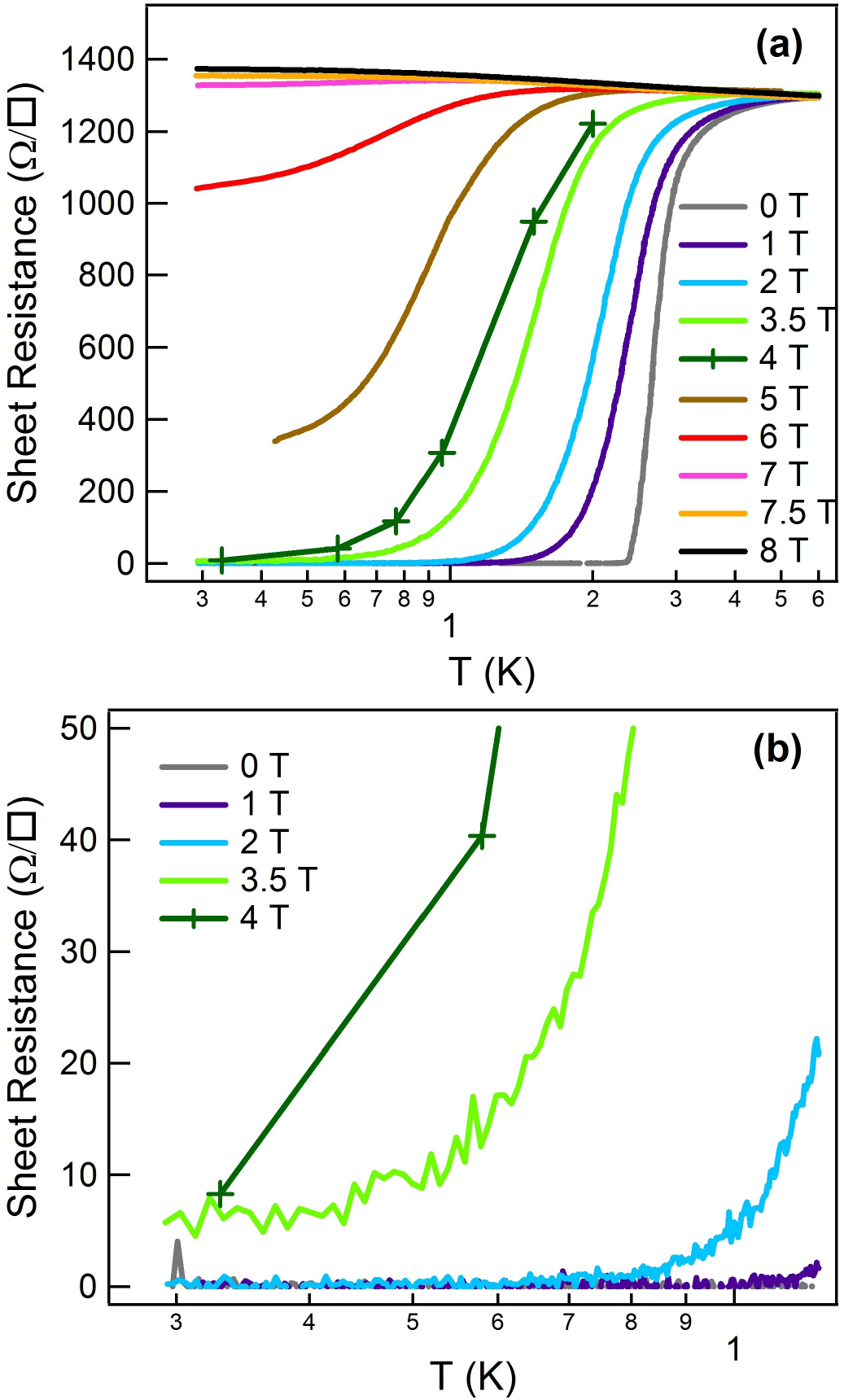}
\caption{(a) Temperature dependence of the $R_{\square}$ measured in the range T = 0.06 - 4 K at fields represented by the color legend. Sheet resistance at 4 Tesla was taken from Fig. 1 (b) in the main text. (b) A close up of the small but finite sheet resistance at fields immediately above $B_{sm}$.}
\label{resistancewith4T}
\end{center}
\end{figure}

The DC sample resistance $R$ measured in the Corbino microwave spectrometer is related to the sheet resistance $R_{\square}$ by the relation $R_{\square} = g R$, where $g = 2\pi/\ln(r_{2}/r_{1})$  is the geometric factor. $r_{2}$ and $r_{1}$ are the outer (2.3 mm) and inner (0.7 mm) radii of the donut shaped sample \cite{scheffler05b,liu11a,liu13a}. Note that the simultaneous measured resistance curve at 4 Tesla for Fig. 1 (a) in the main text is missing due to some experimental problem that corrupted that data set. We took data points at 4 Tesla from magnetic field scans from Fig. 1 (b) in the main text and plot $R_{\square}$ vs $\log T$  together with the simultaneous measured sheet resistance at other fields as shown in Fig. \ref{resistancewith4T} (a) in this supplemental material. The close up of the small but finite sheet resistance at the lowest measured temperatures ($\sim$ 300 mK) for B $>B_{sm}$ is shown in  Fig. \ref{resistancewith4T} (b). The sheet resistance for fields above $B_{sm}$ has a weak dependance on temperature and extrapolate to a finite value at zero temperature. Taking B = 3.5 Tesla as an example, the sheet resistance appears to saturate at approximately 5 $\Omega$.

\begin{figure}[h]
\begin{center}
\includegraphics[width=0.95\columnwidth,angle=0]{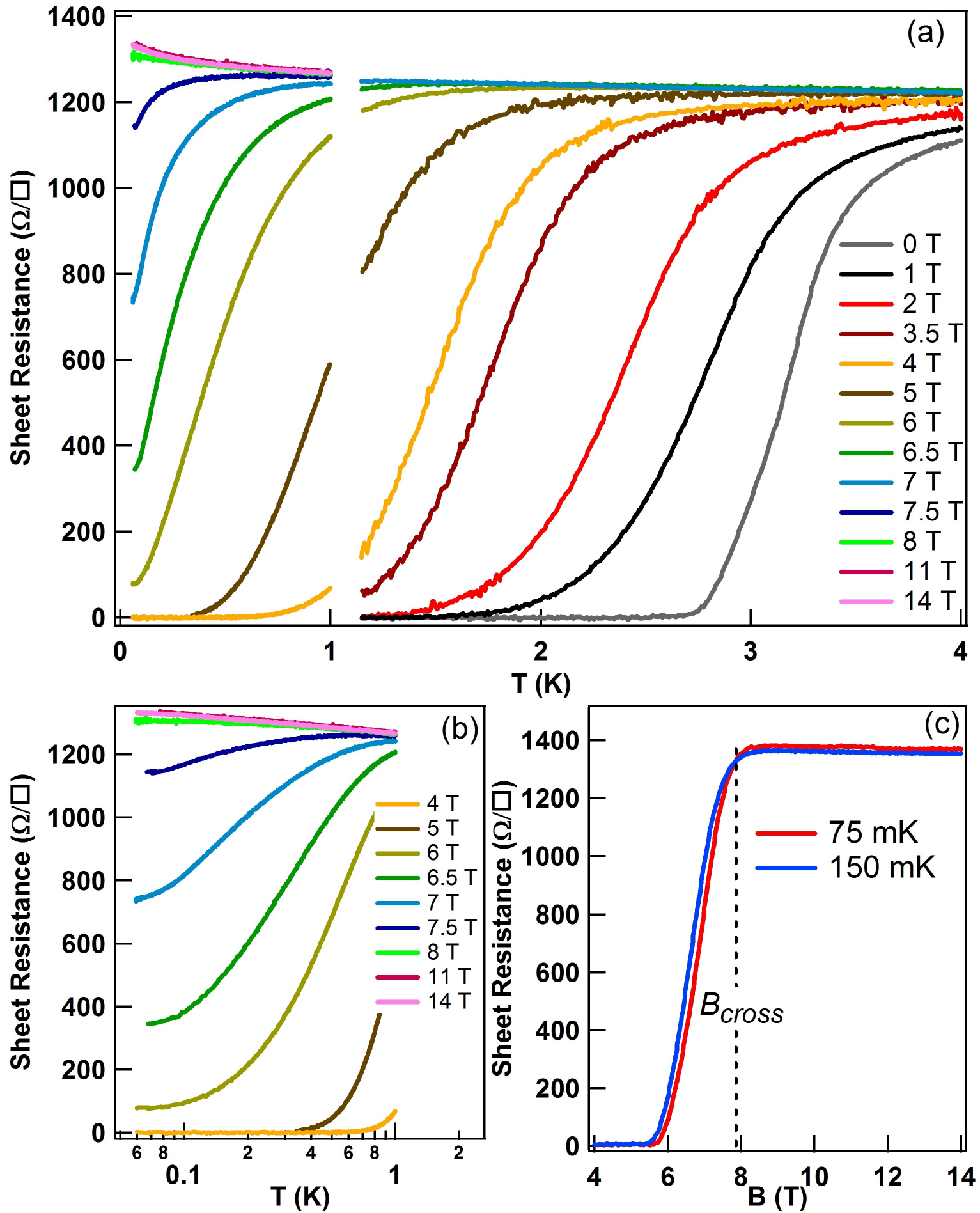}
\caption{(a) Temperature dependence of the $R_{\square}$ measured in the range T = 0.06 - 4 K at fields represented by the color legend.
(b) A close up of  to show the saturation in the sheet resistance. (c) Sheet resistance as a function of field at two fixed low temperatures.}
\label{dillfridgewithzoomout}
\end{center}
\end{figure}

To carefully examine the saturation in resistance, a separate measurement of the same sample down to 60 mK was carried out roughly half a year after the microwave measurements as shown in Fig. \ref{dillfridgewithzoomout} (a). Unfortunately, a direct comparison with the resistance data in the main text could not be made as the InO$_x$ sample anneals even at room temperature over the course of the intervening months. The sample had been kept in a dry box, but presumably lost some oxygen over the course of the intervening months. Although the sample changed with a systematic movement of curves with the same resistance to higher fields, the data sets are overall very similar with continuing trend towards saturation at low temperatures.

In Fig. \ref{dillfridgewithzoomout} (a), we display sheet resistance $R_{\square}$ as a function of temperature at different fields. Despite the lower level of disorder and a new $T_c$ of  2.68 K  at zero field, the sheet resistances at intermediate field range (6, 6.5, 7, and 7.5 Tesla) still become temperature independent approaching T = 0 as clearly shown in the $R_{\square}$ vs $\log T$ plot in Fig. \ref{dillfridgewithzoomout} (b).  We show the sheet resistance as a function of field at 75 mK and 150 mK in Fig. \ref{dillfridgewithzoomout} (c) and one can  see that the $B_{cross}$ changes to 7.86 Tesla.

\section{Methods}
Samples are morphologically homogeneous InO$_x$ films prepared by e-gun evaporation of In$_2$O$_3$ to a thickness of approximately 30 nm onto high-resistivity Si substrates as described elsewhere \cite{MurthyPRL04a}. TEM-diffraction patterns of InO$_x$ films prepared in the same fashion show diffusion rings with no diffraction spots. The featureless AFM image down to a scale of a few nanometers and  the lack of the reentrant behavior in $R$ versus $T$ curves\cite{MurthyPRL04a,CranePRB07b} further support that the films are morphologically homogeneous.  We can estimate the effective disorder level using the standard formulas for transport in 3D to calculate $k_Fl$. $k_F = (3\pi^2 n)^{(1/3)}$ and $l = v_F\tau$, where $v_F= \hbar k_F/m$ is the Fermi velocity. The scattering time $\tau$ can be estimated from the DC conductivity as $1/(R_{\square} d) = n e^2 \tau/m$, where d = 30 nm is the thickness of the sample. The typical carrier density n is on order of $10^{20}$ $cm^{-3}$ \cite{shahar92a,steiner}.

The data reported in this paper were taken in a home build broadband phase sensitive microwave spectrometer \cite{liu13a}. In this technique one measures the complex reflection coefficients $S_{11}^m$  from a thin film sample which requires the use of three calibration standards with known reflection coefficients \cite{Scheffler04a}. The actual reflection coefficient at the sample surface $S_{11}^a$ is expressed as $S_{11}^a = \frac{S_{11}^m - E_D }{E_R + E_S (S_{11}^m - E_D)}$. Here, the complex error coefficients  $E_D$, $E_S$ and $E_R$ represent the effects of extraneous reflections, phase shifts and losses in the coaxial cables. They are all temperature, frequency and field dependent. The sample sheet impedance $Z_S^{eff}= g \frac{1+S_{11}^a}{1-S_{11}^a}Z_{0}$, where  $Z_{0}=$50 $\Omega$. In the thin film limit, complex sheet conductance $G \equiv \sigma d $ is related to sheet impedance as $G = 1/Z_s$. The contribution of the Si substrate to the sample impedance is obtained by comparing the measured zero field AC signal of the InOx/substrate system in the normal state with the normal-state frequency independent InO$_x$ film impedance value. See Ref. \cite{BoothPRL96a,liu11a} for more detailed information.

In current study, we use 20nm NiCr on Si, a blank high-resistivity Si substrate and a bulk copper sample as the three calibration standards.  We calibrated the spectrometer at each displayed magnetic field except 4 Tesla due to a missing set of calibration curves. The data at 4 Tesla were calibrated using an effective calibration  interpolated from 3.5 Tesla and 5 Tesla calibration standards.  We estimate that the error introduced by this interpolation is less than 1 \% \cite{liu13a}. The sample at 5 Tesla had a impedance closely matched to the cable thus  had a maximum absorption of microwave radiations. The lowest temperature for that field  was 426 mK under microwave radiation. The heating effects at other fields were negligible.

In the previous publication, we used a thick Nb film as a ``short" standard to calibrate the spectrometer at low temperatures \cite{liu11a}. A superconducting Nb film as such is one of the best calibration short standards to measure a disordered superconductor more accurately. However, for measurements and calibrations in magnetic fields we cannot use a superconductor as a short standard since we calibrated the spectrometer at each displayed magnetic field unless otherwise specified. Therefore we have used a polished bulk copper piece with thick gold film on top as a short standard instead. Although it works fine for most of the finite field measurements, the choice of the bulk copper does not work very well for the zero field measurements because InO$_x$ at low temperatures in zero field has an impedance many orders of magnitude different than a bulk copper standard. However, at finite magnetic fields, InO$_x$ becomes dissipative enough that it has a impedance much closer to copper than the zero field data.

\begin{figure}[h]
\begin{center}
\includegraphics[width=\columnwidth,angle=0]{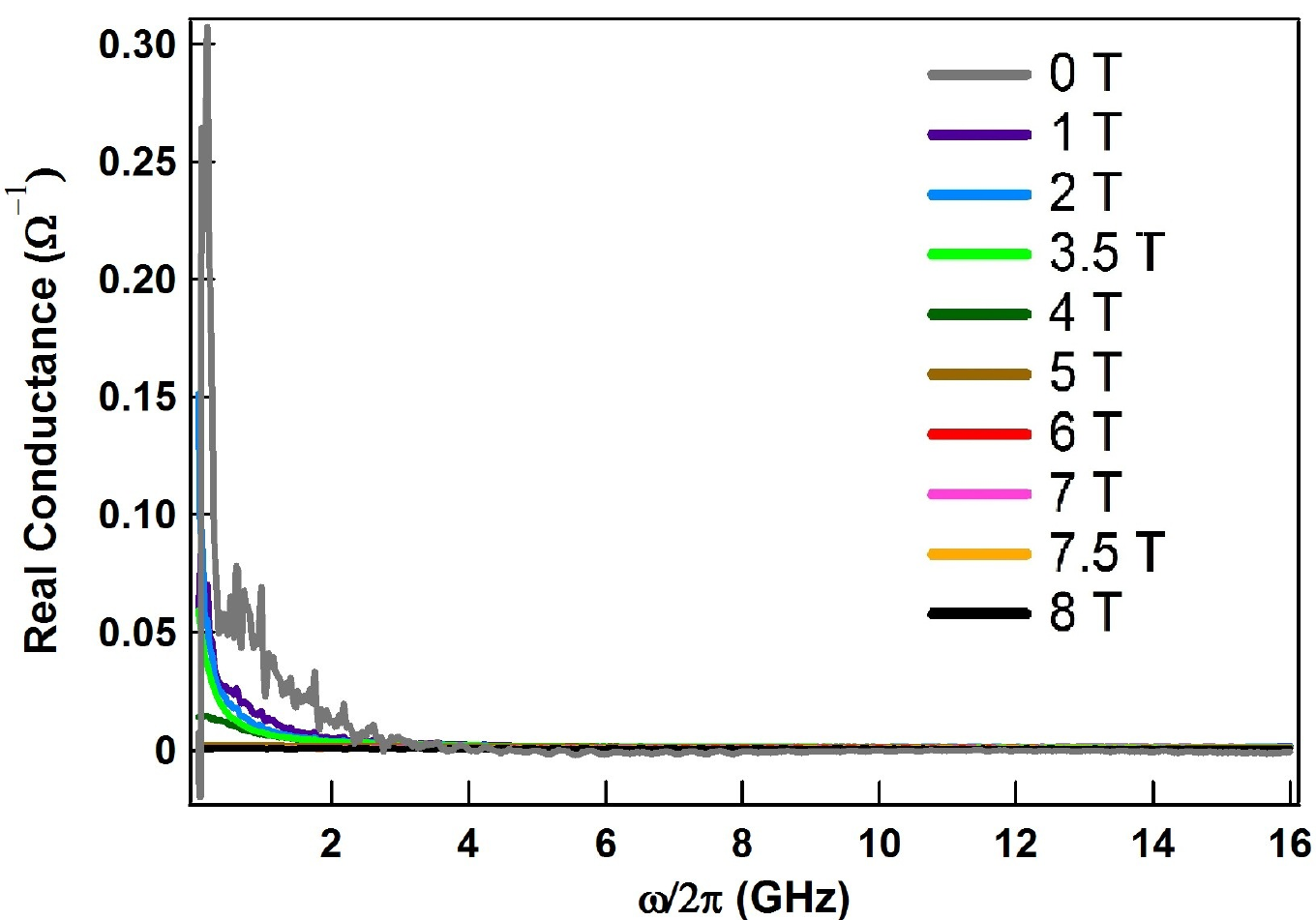}
\caption{The real conductance ($G_1$)  as a function of frequency at the base temperature for each field.}
\label{condinlinearplot}
\end{center}
\end{figure}

In Fig. \ref{condinlinearplot}, we plot the real conductance $G_1$ as a function of frequency at the base temperature for each field on a linear scale.  As we stated in the main text, $G_1$ at zero fields stretches down to noise level at higher frequencies. At some frequencies, $G_1$ at zero fields becomes slightly negative showing the limit of our spectrometer.   At  low frequencies, $G_1$ becomes slightly positive with substantial noise.   We believe all these features are due to errors in establishing the proper phase reference by using  copper as a  short standard when one tries to measure a  very low impedance superconducting film.

One also can estimate the contribution from the quasiparticles to $T_{\theta}$ for B = 0 at low temperatures. Since the  typical carrier density n is on order of $10^{20}$ $cm^{-3}$ \cite{shahar92a,steiner}, from the DC resistivity one can estimate that the mean free path $l \approx 2$ nm and the normal state scattering rate is about 100 THz. Using $T_{\theta}[T \rightarrow 0] \approx$ 34 K, we can estimate the maximum contamination contribution from the quasiparticles to the superfluid stiffness as $T_{\theta, quasi} = T_{\theta}[T \rightarrow 0]\frac{(\omega\tau)^2}{1+(\omega\tau)^2}\approx 3.4 \times 10^{-7}$ K at 10 GHz.   This is the amount of contamination contribution to the superfluid stiffness if all the superconducting electrons become subject to normal state dissipative processes.  Although this is obviously an overestimate, it still is a very small number and below our detection limit.

\section{Fits of the AC sheet conductance}
\begin{figure}[h]
\begin{center}
\includegraphics[width=\columnwidth,angle=0]{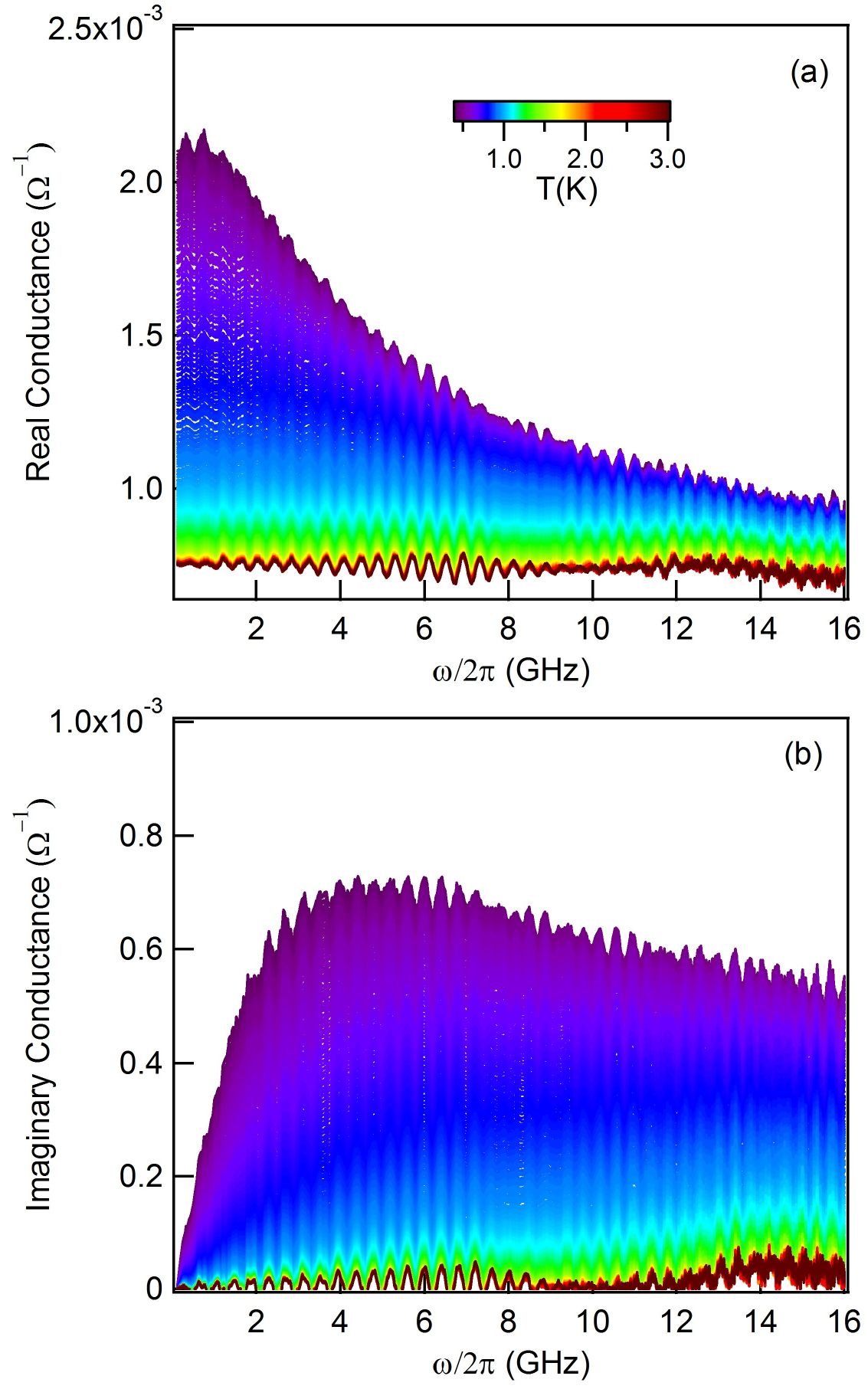}
\caption{The (a) real ($G_1$) and (b) imaginary ($G_2$) conductance respectively as a function of frequency at different temperatures  for B = 5 Tesla.}
\label{Conductanceat5T}
\end{center}
\end{figure}

In Fig. \ref{Conductanceat5T}, we plot the complex sheet conductance as a function of frequency at different temperatures at 5 Tesla. Data at other fields are similar. One  observes a finite width in $G_1$  approximately 4 GHz  at the lowest temperature. $G_2$ also feature a maximum at about the same frequency. This frequency is much smaller than the estimated normal state scattering rate, which is on order of a hundred THz for a disordered film like InO$_x$. We interpret the frequency where the peak in $G_2$ is located as the fluctuation frequency $\Omega$ of the Cooper pairs. As we discussed in the main text, we fit $G_1$ and $G_2$ to a Lorentzian lineshape:
\begin{equation}
G = \sigma d = \frac {n_n e^2 d\tau_n}{m} \frac{1}{1-i\omega\tau_n}+ \frac {n_s e^2 d\tau_s}{m} \frac{1}{1-i\omega\tau_s}.
\label{fittwodrude}
\end{equation}
We use two Drude terms to describe the combined contributions from the normal electrons and superconducting fluctuations to the complex conductance. The use of the Drude model follows from an assumption of exponential relaxation of charge currents in time. $d$ is the thickness of the sample. $1/\tau_n$ is the scattering rate of normal electrons and $1/\tau_s$ corresponds to the superconducting fluctuation rate $\Omega$ in the main text. $n_n$ and $n_s$ are the number density of normal electrons and superfluid density respectively.

\begin{figure}[h]
\begin{center}
\includegraphics[width=\columnwidth,angle=0]{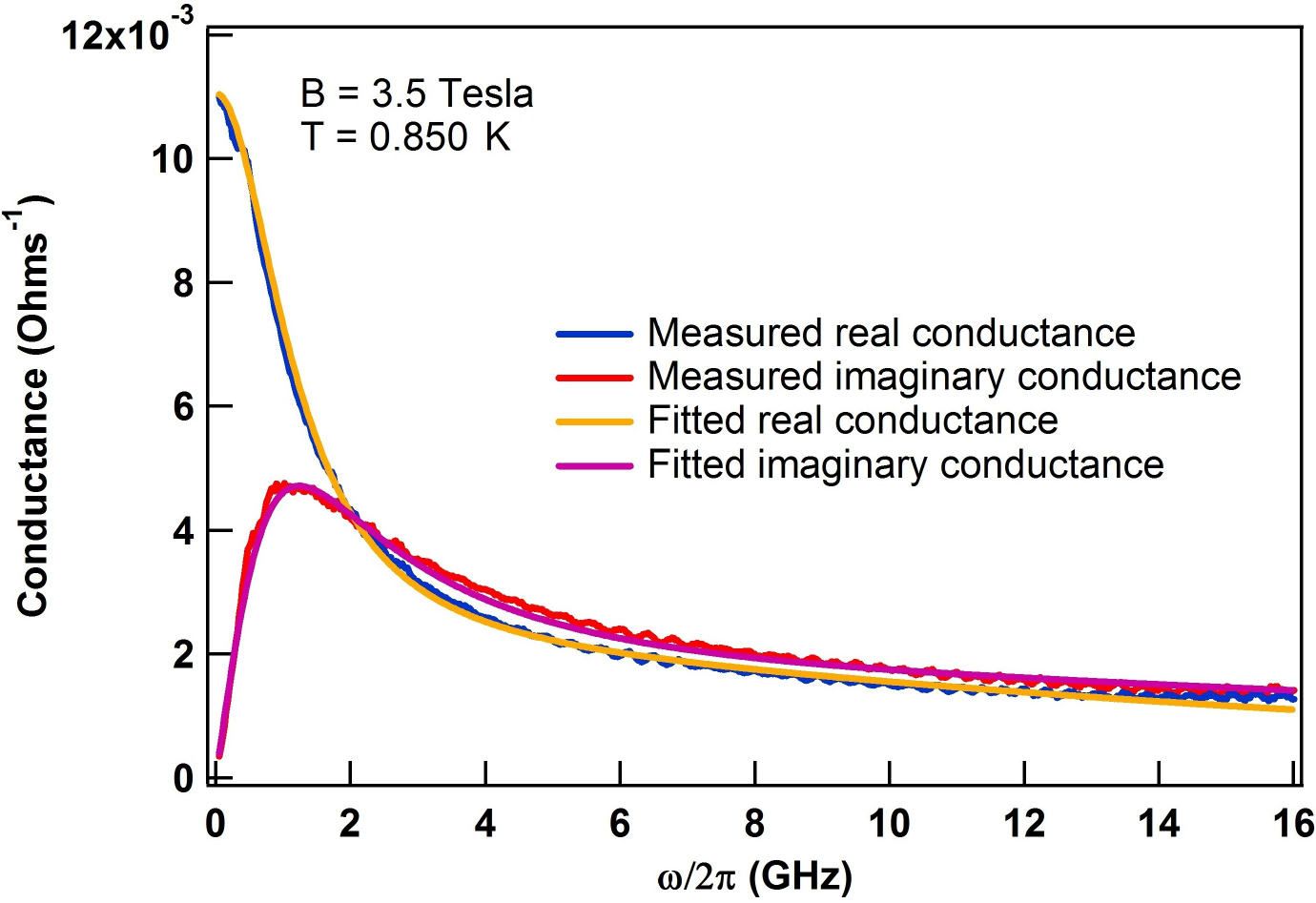}
\caption{Experiment measured $G_1$ and $G_2$ and fitted data to a Lorentzian lineshape model.}
\label{fitting}
\end{center}
\end{figure}

The scattering rates of the normal electrons for this film are in general orders of magnitude bigger than $\Omega$ and exceed our frequency range. Therefore, the Drude term of the normal electrons just makes a constant  contribution to $G_1$ and a negligible contribution to $G_2$. The frequency dependence in the real and imaginary conductance we observe in Fig. \ref{Conductanceat5T} is from the contribution of the fluctuating Cooper pairs. Therefore, in this model the fitted width of $G_1$ is the characteristic fluctuation rate $\Omega$(T)$=1/\tau_s$, while its integrated area is equivalent to the high frequency limit of superfluid stiffness $T_{\theta} = \frac{n_sh^2d}{8\pi m k_B}$ (in appropriate units). As the temperature is increased, we see that the response from Cooper pairs vanishes. At high enough temperatures, we return to the AC response for an ordinary disordered metal that is associated with normal electrons. Under the assumption $1/\tau_n \gg 1/\tau_s$, within our accessible frequency range and also within the frequency range where the peaks in $G_2$ locate, Eq. \ref{fittwodrude} can be reduced to:
\begin{equation}
G = \sigma d = G_n + \frac {n_s e^2 d\tau_s}{m} \frac{1}{1-i\omega\tau_s}
\label{fitdrudeandconst}
\end{equation}

\noindent where $G_n$ is a constant and is the contribution from normal electrons. Since $T_{\theta}$ is defined as $k_B T_{\theta} = \frac{G_2}{G_Q} \hbar \omega = \frac{N_s(\omega)e^2\hbar d}{mG_Q}$, we have $N_s(\omega) = \frac{n_s\omega^2\tau_s^2}{1+\omega^2\tau_s^2}$. In Fig. \ref{fitting}, we show the fitting to $G_1$ and $G_2$ at B = 3.5 Tesla, T = 850 mK as an example. The data is fit very well by the proposed functional form.
For this particular data set, in the low frequency regime, the complex conductance can be written as,
\begin{equation}
G  = \left(0.0017 + \frac{0.0094}{1-i\omega/(2\pi \times 1.239) GHz}\right)\ \ \Omega ^{-1}.
\end{equation}
We would like to emphasize here that the fact that the data is fitable by the Drude model only means that the spectra is Lorentzian and time correlations are exponential. It does not mean that the conductivity is due to electrons

\section{Relevance of the Kosterlitz-Thouless-Berezinskii universal jump}

In the conventional wisdom \cite{fisher90a}, the KTB transition induced by unbinding of thermally generated
vortex-antivortex pairs only occurs at B = 0.  In the zero disorder limit and at finite field, the field- or temperature-tuned transition occurs through unbinding of dislocation-antidislocation pairs and melting of the Abrikosov lattice.  Therefore, at finite disorder and finite field, no phase transition is expected at finite temperature.  Even an infinitesimal amount of thermal energy should result in some vortex motion as there will be rare regions where the activation energy vanishes.

\begin{figure}[h]
\begin{center}
\includegraphics[width=\columnwidth,angle=0]{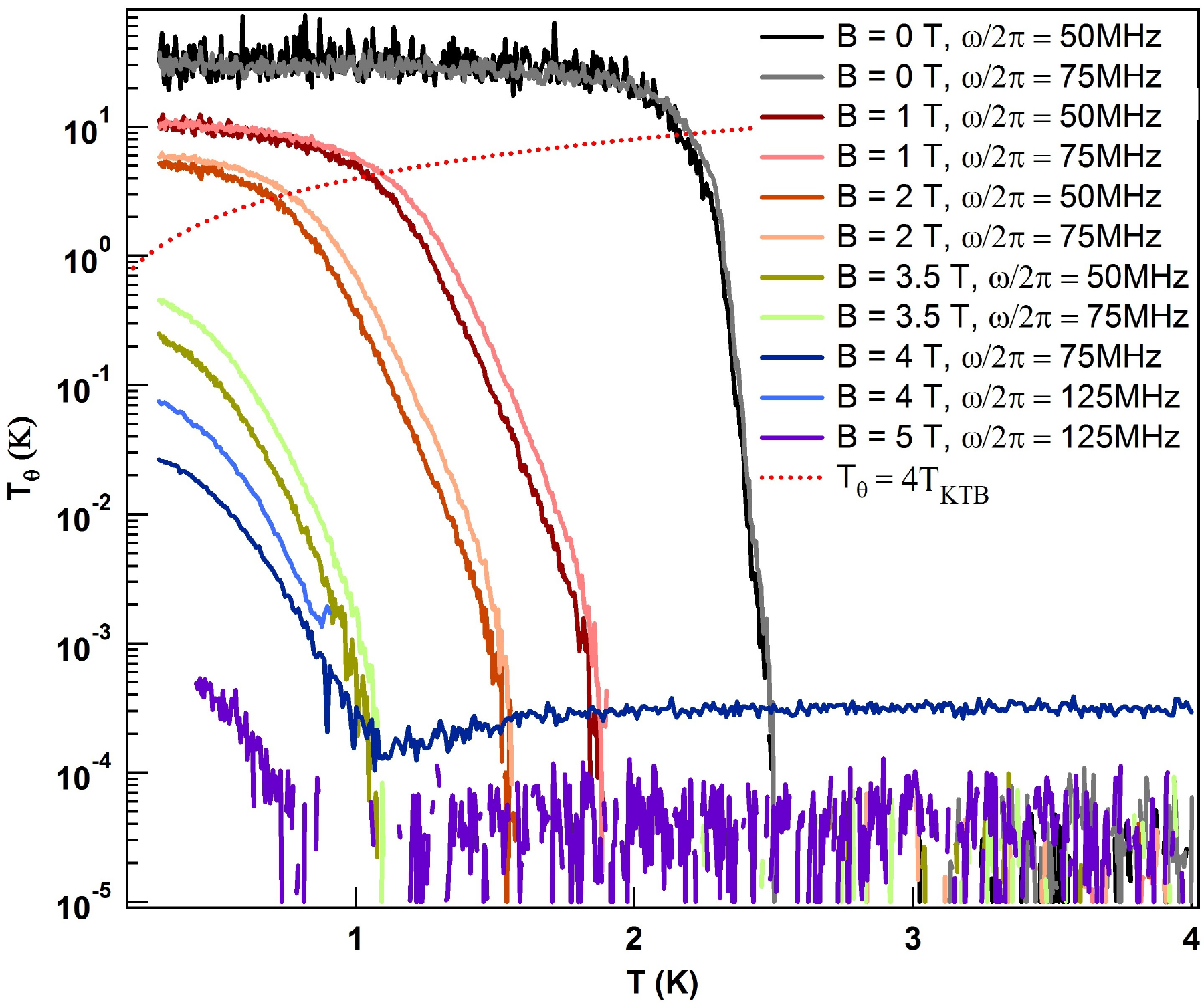}
\caption{Temperature dependence of superfluid stiffness at several representing magnetic fields and frequencies. The dashed line is the KTB prediction for the universal jump in superfluid stiffness. Here we show that the relevance of the universal predication line $T_{\theta} = 4 T_{KTB}$ even in the presence of an applied magnetic field for field below $B_{sm}$. }
\label{universaljump}
\end{center}
\end{figure}

Recently however, Misra $et$ $al.$ \cite{misra13a} observed a transition in finite field with features reminiscent of a KTB transition using a low frequency mutual inductance technique.   We observe similar phenomena in the present data measured at much higher frequencies.  The superfluid stiffness at several representative fields and the lowest measured frequencies is plotted as a function of temperature in Fig. \ref{universaljump}. The universal jump prediction $T_{\theta} = 4 T_{KTB}$ crosses the stiffness curves very close to where they start to spread for fields below $B_{sm}$.  This indicates the relevance of some aspect of KTB physics, although it difficult to conclude definitely from this set of data (and the data of Misra $et$ $al.$) whether or not a true phase transition is occurring or whether the system is just experiencing a strong crossover at a temperature set by the phase stiffness.   Above $B_{sm}$, the universal prediction seems to be irrelevant to the superfluid stiffness curves, indicating that the sample transitions out of the superconducting state.

\bibliography{ScIns}

\end{document}